\documentclass{ws-procs9x6}

\begin{document}

\title{Gluonic Meson Production
%\footnote{\uppercase{T}his work is supported by etc, etc.}
}
\author{PETER MINKOWSKI\footnote{\uppercase{W}ork partially
supported by grant 200020-100122 of the \uppercase{S}wiss 
\uppercase{N}ational \uppercase{S}cience \uppercase{F}oundation.}}

\address{Institute for Theoretical Physics, Univ. of Bern, \\
CH-3012 Bern, Switzerland\\
E-mail: mink@itp.unibe.ch}

\author{WOLFGANG OCHS}

\address{Max-Planck-Institut f\"ur Physik (Werner Heisenberg-Institut) \\
F\"ohringer Ring 6 \\
D-80805 M\"unchen, Germany\\
E-mail: wwo@mppmu.mpg.de}

\maketitle

\abstracts{
The existence of glueballs is predicted in QCD, the lightest one
with quantum numbers $J^{PC}=0^{++}$, but different calculations do not
well agree on its mass in the range below 1800 MeV.
Several theoretical schemes have been proposed to cope with the 
experimental data which often have considerable uncertainties. 
Further experimental studies of the
scalar meson sector are therefore important and we discuss recent proposals
to study leading clusters in
gluon jets and charmless $B$-decays to serve this purpose.
}

\section{Introduction}
The existence of glueballs, i.e. bound states of two or more gluons, are
among the early predictions of QCD. There is general agreement
in that
the lightest glueball should be in the scalar channel with $J^{PC}=0^{++}$.
The experimental situation is still controversial as the properties of
scalar resonances are not well known. This can be illustrated by
looking at the Particle Data Group\cite{pdg} listing 
on the lightest scalar mesons  (status end 2003).
Glueballs can be searched for among the 5 isoscalar states with mass below
1800 MeV, listed in Table \ref{tab:scalars}. The nature of two broad states 
$f_0(600)$ (also called $\sigma$) and $f_0(1370)$ is still controversial.
Whereas quite a number of decay modes
are ``seen'', no branching ratios are quoted for any one of these particles.
Only few ratios of rates are considered as acceptable of which altogether
only two have been measured by two experiments independently.

\begin{table}[t]
\tbl{Information on isoscalar mesons with $J^{PC}=0^{++}$ 
from the Particle Data Group. \vspace*{1pt}  }
{\footnotesize
\begin{tabular}{|l|ccccc|}
\hline
\text{states} & $f_0(600)$ & $f_0(980)$& $ f_0(1370)$& $
f_0(1500)$&$f_0(1710)$ \\
\text{width [MeV]}& 600-1000 & 40-100 &300-500 & $109\pm 7$ & $125\pm 10$ \\
\hline
 \text{hadronic modes ``seen''} & 1 & 2 & 11 & 13 & 3 \\
\text{ratios of rates published} & - & 1 & 13 & 18 & 5 \\
\text{``used'' by PDG} & - & 0 & 0 & 5 & 2 \\
\text{from more than 1 exp.} & - & 0 & 0 & 2 & 0 \\
 \hline 
  \end{tabular}
 \label{tab:scalars}  }
\vspace*{-13pt}
\end{table}     

Besides the decay branching ratios the production properties are important
for the identification of gluonic mesons. Glueballs should be produced with
enhanced rate in a ``gluonic environment'', such as central hadronic
production (double Pomeron exchange), radiative $J/\psi$ decay, proton
antiproton annihilation, decay of excited Quarkonia into their respective 
ground states,
whereas their production should be suppressed in $\gamma\gamma$ collisions.

There is not yet an agreement on the mass of the lightest $0^{++}$ glueball 
and its mixing with other states, it is even debated whether it has been
seen at all. It is rather clear that 
the evidence for the scalar glueball will emerge only from a thorough
experimental study of the low mass scalar channels and the identification of
the scalar $q\bar q$ nonet in a parallel effort. Various channels should be
probed with the aim to obtain more precise information on 
the production and decay of the scalar resonances.

A recent analysis of $K\bar K$ mass spectra at HERA\cite{zeus} 
with a prominant peak near the mass of 1700 MeV (presumably $f_0(1710)$) 
has shown the potential for hadron spectroscopy at HERA.
In this report we will discuss the recent proposal to search for glueballs
in the leading part of the gluon jet\cite{mo} and first results\cite{bm} 
(see also discussions in%
\cite{pw,rs,sz}), which may be applicable at HERA. Furthermore we report
on the possibility for glueball searches and results in $B$ decays.\cite{mo1}

\section{QCD Expectations}
The basic triplet of binary glueball states which can be formed by two
``constitutent gluons'' corresponds to the three invariants which can be
built from the bilinear expressions of gluon fields and carry quantum numbers
$J^{PC}=0^{++},\ 0^{-+}$ and $2^{++}$.\cite{fm} 
Quantitative results are derived today from the QCD lattice calculations or
QCD sum rules, both agree that the lightest glueball has quantum numbers
 $J^{PC}=0^{++}$.

Lattice calculations in 
quenched approximation%
\cite{bali,svw,Morning,lt} (without light sea quark-antiquark pairs)
suggest
the lightest glueball to have a mass in the range 1400-1800 MeV.%
\cite{balirev} Results from unquenched calculations still suffer from 
systematic effects, the large quark masses of the order of the strange 
quark mass and large lattice spacings. Typically, present results
on the glueball mass are about 20\% lower 
than the quenched results.\cite{ht,hnm}
Another interesting result would be the mass of the light scalar $\bar q q$
mesons. A recent result\cite{hnm} suggests the mass ($1.0\pm 0.2$) GeV 
for the 
scalar $a_0$; this is well consistent with the mass of $a_0(980)$ but in
view of the systematic uncertainties it is not yet possible to exclude 
$a_0(1450)$ as the lightest isovector scalar particle.

Results on glueballs have also been obtained from QCD sum rules. 
Recent calculations\cite{narison} for the  $0^{++}$ glueball
yield a mass consistent with the quenched lattice result
%(with upper bound 2.16 $\pm$ 0.22 GeV), 
but in addition require a gluonic state near 1 GeV.
A strong mixing with  $q\overline q$ is suggested 
resulting finally in the broad $\sigma$ and narrow $f_0(980)$.
Similar results with a low glueball mass around 1 GeV are obtained 
also in other calculations.\cite{steele} On the other hand, it is
argued\cite{forkel} that the sum rules
can also be saturated by a single glueball
state with mass $1.25\pm 0.2$ GeV. 

In conclusion, there is agreement in the QCD based calculations on the
existence of a $0^{++}$ glueball but the mass and width of the lightest state 
is not yet certain 
and phenomenological searches should allow a mass range of about
1000-1800 MeV.  

There is also another class of gluonic mesons, the hybrid $q\bar q g$ 
states, both in the theoretical and
experimental analysis, but we will not further consider these here (see, for
example, review\cite{klempt}).

\section{Spectroscopy of Scalar Mesons - Phenomenology}

In the mass range below 1800 MeV the PDG lists two isovectors
$a_0(980)$ and $a_0(1450)$ as well as the strange $K^*_0(1430)$, 
furthermore, there is a possibility
of a light strange particle $\kappa(800)$. From these particles 
and the isoscalars in Tab. \ref{tab:scalars} one should build the relevant
multiplets, one (or more) $q\bar q$ nonets and a glueball. There are various
schemes for the spectroscopy of the light scalars. We emphasize two different 
routes in the interpretation of the data, 
which are essentially different in the classification of the states,
both with some further possibilities in details.

\noindent {\it Route I: ``Heavy'' $q\bar q$ multiplet (above $f_0(980)$) and
``heavy'' glueball}\\
One may start from the glueball assuming a mass
around 1600 MeV as found in
the quenched lattice calculations emphasized above.
In the isoscalar channel
there are the states  $f_0(1370),\ f_0(1500)$ and $f_0(1710)$ 
nearby in mass which are assumed to be
mixtures of the two members of the nonet and the glueball.
The multiplet of higher mass then includes furthermore 
the uncontroversial $q\bar q$ state
$K^*(1430)$ and also the nearby $a_0(1450)$.
After the original proposal\cite{ac} several such mixing schemes have been
considered (review\cite{klempt}) using different phenomenological
constraints. There are schemes\cite{ac} 
with the largest gluon component residing in
$f_0(1500)$ and others\cite{svw} where this role is taken by $f_0(1710)$.
In these schemes $f_0(980)$ and $ a_0(980)$ are 
sometimes superfluous, they are taken as 
multiquark boundstates\cite{Jaffe} or $K\overline K$
molecules\cite{weinstein} and then are removed from $q\bar q$ spectroscopy.
An attractive possibility is the existence of an additional light nonet,
which includes $\sigma/f_0(600)$, $\kappa$, $a_0(980)$ and $f_0(980)$
either as $q\bar
q$ or of $qq\bar q\bar q$ bound states. Such schemes appear in
theories of meson meson scattering in a realization of chiral symmetry,
for an outline, see review\cite{ctoe}. 

\noindent {\it Route II: $q\bar q$ multiplet including $f_0(980)$ and ``light''
glueball}\\
Alternatively, one may start from an identification of the lightest nonet
and then look for the glueball among the remaining states.\cite{mo0}
Several approaches agree on a similar nonet with $f_0(980)$ as the lightest 
member,\cite{instanton,mo0,anis,narison} although with different intrinsic
structure. Also $K^*(1430)$ belongs to this nonet whereas the identification of the other
members differs. There are arguments\cite{mo0,instanton} 
for a strong flavour mixing 
 similar to the pseudoscalar sector with the correspondence
$f_0(980)\leftrightarrow \eta'$ near flavour singlet and 
$f_0(1500)\leftrightarrow \eta$ near flavour octet. The isovector could be
$a_0(980)$\cite{mo0} or $a_0(1450)$.\cite{instanton} There is no room for
$\kappa$ in these schemes.

The remaining light scalars, the broad $\sigma$ and $f_0(1370)$,
are then candidates for the lightest glueball. The interpretation
of $\sigma$ which is related to the strong $\pi\pi$ scattering up to
 1 GeV is subject to intense discussions and
controversies.\cite{woaschaff} 
In our phenomenological analysis\cite{mo0} we consider
both states as a single object with a width of
about 500-1000 MeV. 

Our arguments in favour of the 
glueball hypothesis include: the strong central production in $pp$
collisions, the appearence in the Quarkonium decays $\psi',\psi''\to
J/\psi\pi\pi$, also in corresponding $Y',Y''$ decays, in $p\bar p$
annihilation 
 and the suppression in $\gamma\gamma$ collisions;\cite{mo2}
on the other hand, contrary to expectation, there is no strong signal in
radiative $J/\psi$ decays.
A glueball in this mass region is also located in the
sum rule analyses\cite{narison,steele,forkel} and the K matrix fits to a
variety of production 
processes.\cite{anis} Alternatively, this
broad $\pi\pi$ ``background'' has been viewed as due to non-gluonic exchange
processes.\cite{klempt}

 As a common feature of the above schemes the lightest
glueball is not expected to appear as a single narrow resonance
but rather as a phenomenon spread over a mass range of around 500 MeV.
Either it is mixed with several moderately narrow 
isoscalar resonances in a range from
1300-1800 MeV (route I) or it appears mainly as a broad state with a 
large width by itself (route II). 
In view of these different possibilities it is important to 
improve our knowledge on gluonic interactions and to look for further
possibilities of glueball production.
\section{Gluonic Meson Production in Gluon Jets} 
New information on gluonic mesons can be obtained from the comparative study
of the leading particle systems in quark and gluon jets.
The possible appearence of isoscalar particles 
in the leading system of a gluon jet has already been considered
long ago.\cite{pw} More specifically, 
the production of glueballs in the fragmentation
region at large Feynman $x$ has been considered.\cite{rs}
The search for glueballs applying a rapidity gap selection of events
 and charge distributions in quark and gluon jets has been suggested
recently.\cite{mo}

There is a well established fragmentation phenomenology for quark jets:
a particle which carries the primary quark  
as valence quark is produced with larger probability
 at high momentum fraction $x$ than other particles.
For example, the $u$-quark
will produce more $\pi^+$ than $\pi^-$ at large $x$.
A natural extension of this phenomenology applies
for gluon jets: particles with large Feynman $x$ are predominantly those
which carry the initial gluon as valence gluon, these are
 glueballs or hybrids if they exist.

\begin{figure*}[ht]
%\centerline{\epsfxsize=3.9in\epsfbox{procs-fig1.eps}}
\centerline{\epsfxsize=5cm\epsfbox{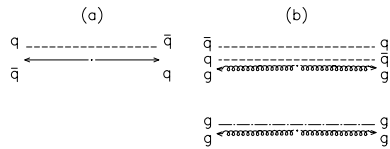}}
%\mbox{\epsfig{file=ochs.fig1,width=10cm}}
%\vspace{-0.7cm}
\caption[]{ 
The colour neutralization (a) of an initial $q\overline q$ pair
by $\overline q q$ and (b) of an initial $gg$ pair by either  double
colour triplet
$q\overline q$ or by colour octet $gg$.}
\label{fig:colour}
\end{figure*}

Whether the idea can be transfered from quark to gluon jets 
in this way depends on
the hadronization mechanism at the distances of about 1 fm, 
where the colour confinement forces become important,
but there is no firm approach to deal with
these non-perturbative processes. 
At the end of a parton cascade the valence 
quark will form a hadron by recombining with an anti-quark corresponding to
an interaction between the colour triplets. In a simple case a primary
pair of energetic $q$ and $\bar q$  
in a colour singlet state is neutralised by a soft 
$\bar q q$ pair, see Fig. \ref{fig:colour}a.
For gluons there are in general two different possibilities.
A pair of energetic valence gluons can be colour neutralized again by
colour triplet forces (two soft $q\bar q$ pairs) or by colour octet forces 
(a pair of soft gluons), see Fig.  \ref{fig:colour}b. Whereas the standard
hadronization models support only the colour triplet neutralization, the 
second mechanism would allow glueball production.

It is not obvious to what extent the two types of neutralization mechanisms
are realized in a given
process. The octet neutralization could in general be a rare process
enhanced only for particular kinematic configurations.
We argued\cite{mo} that in a gluon jet with a large rapidity gap the octet mechanism
should become visible if it exists. For large rapidity gaps
the more complex multi-quark pair exchanges through the gap will be suppressed. 
 
The relative importance of the colour octet mechanism can be tested
by studying the distribution  of 
electric charge of the leading hadronic cluster beyond
the gap. These charges should approach for large gaps a limiting
distribution corresponding to the minimal number of partons traversing the
gap for the considered neutralization mechanism.

There is a clear distinction between the triplet and octet mechanisms: the
leading charge in the minimal triplet ($q\bar q$) configuration is $Q=0,\pm1$
whereas for the octet configuration ($gg$) it is $Q=0$. Note that this
result is independent of the existence of glueballs. On the other hand, if no
extra $Q=0$ component is observed, then there is no evidence for the 
octet mechanism whose existence is 
a precondition for the production of glueballs: glueballs could exist
in nature but  not be produced in gluon jets. 

\begin{figure}[t]
\begin{minipage}{5cm}
\centerline{\epsfxsize=6cm\epsfbox{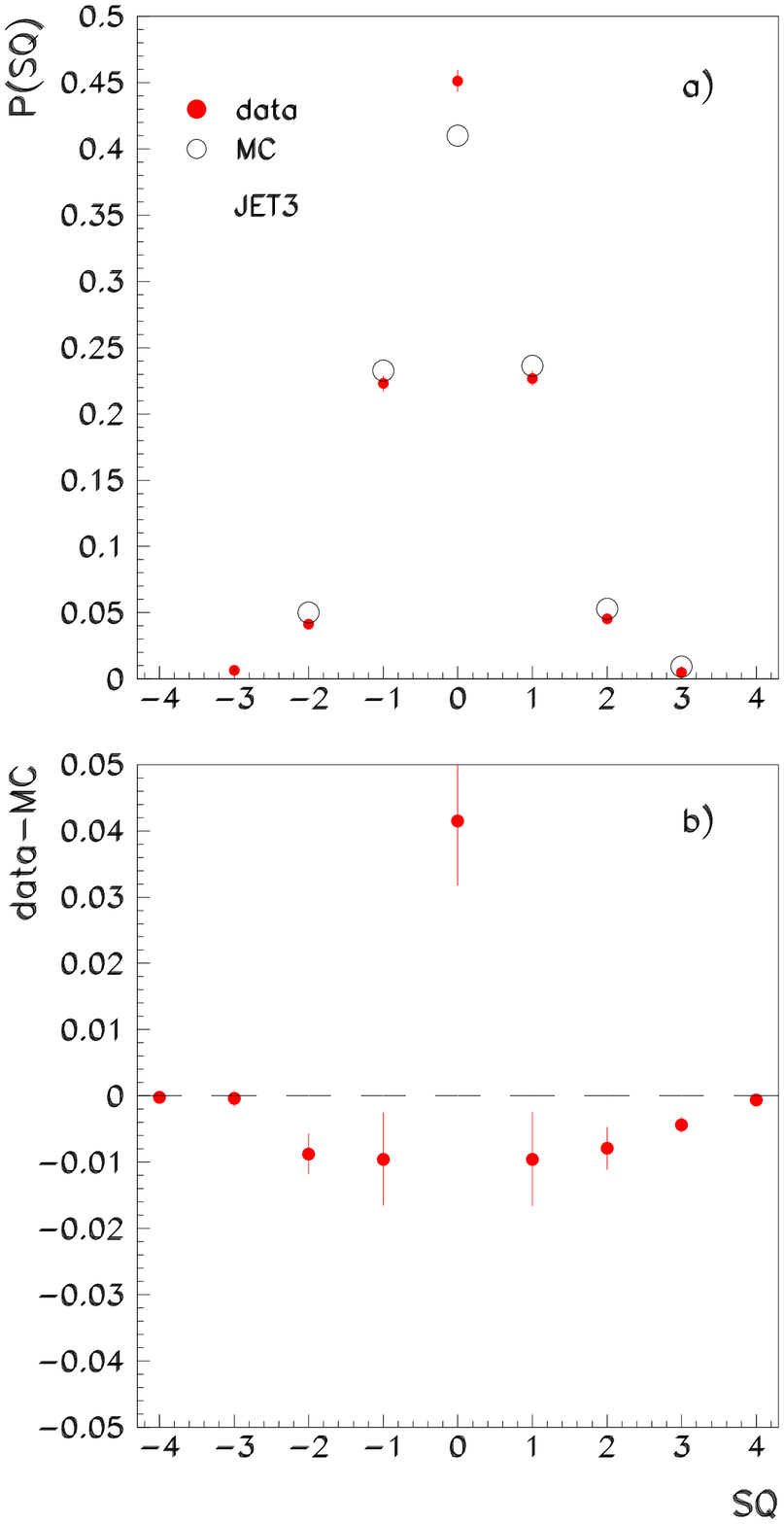}}
\end{minipage} 
\hfill
\begin{minipage}{5cm}
\centerline{\epsfxsize=6cm\epsfbox{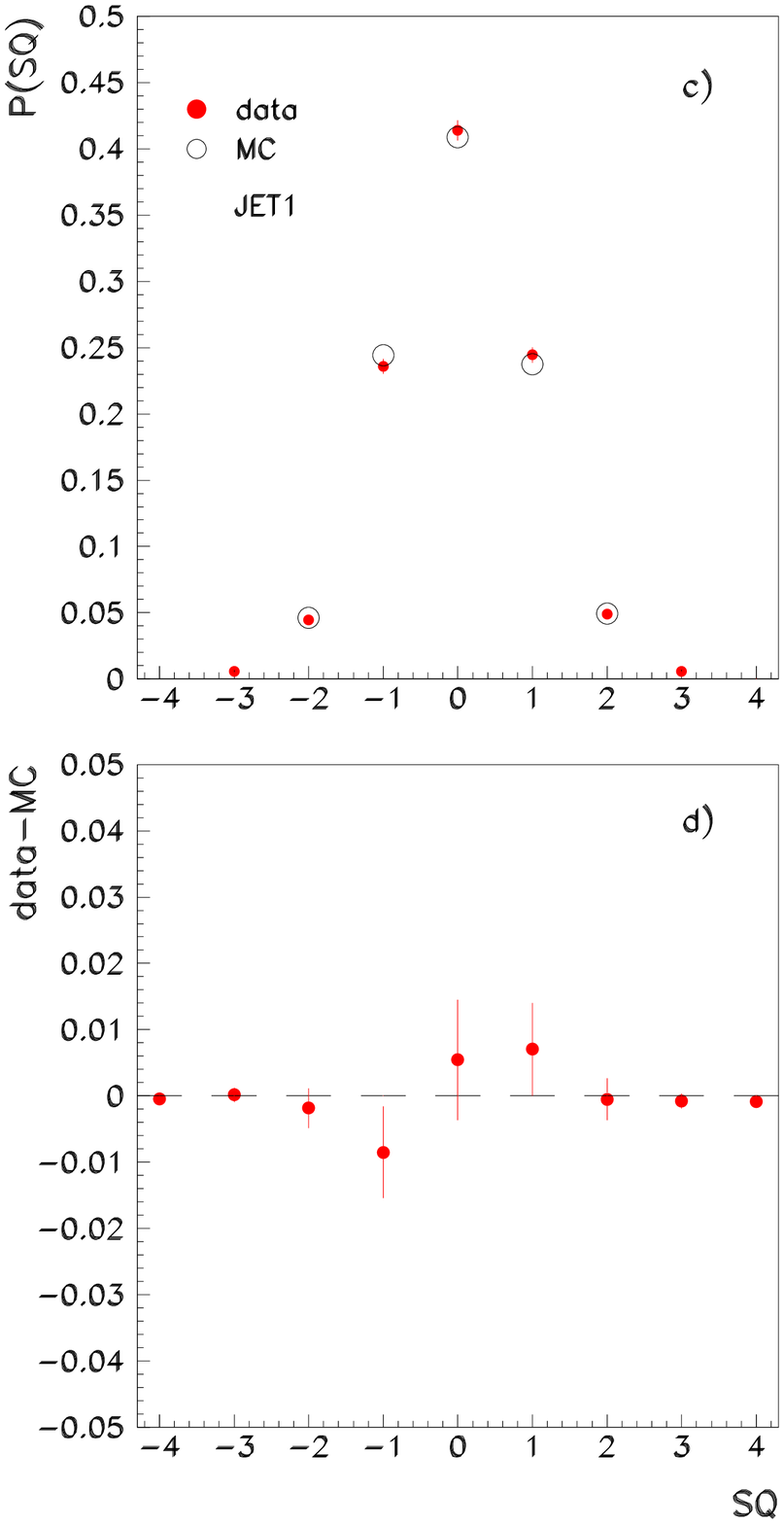}}    
\end{minipage}
\vspace{0.5cm} 
\caption{
Sum of charges of leading system for a,b) gluon jets and c,d) quark jets
(preliminary results 
by DELPHI\protect\cite{bm}) showing the significant excess
of data for gluon jets over the expectations from the JETSET MC.}
\label{fig:charges}
\end{figure}

An exploration of these possibilities has been performed in $e^+e^-$
annihilation at LEP, first with the DELPHI data\cite{bm} and recently also
with OPAL\cite{opal} and ALEPH data.\cite{aleph} For illustration we show
in Fig.~\ref{fig:charges} the results by DELPHI.
The charge distribution
has been obtained for the leading cluster, as defined by a rapidity gap
$\Delta y=2$, in
quark and gluon jets. The data with this selection
do not yet show the limiting charge distribution 
as can be concluded from the presence of charges 
$Q=\pm2,\pm3$ at the level of 10\%. 
The data for quark jets are in a good agreement with 
the JETSET Monte Carlo\cite{jetset} which is based on
the triplet neutralization mechanism 
whereas in gluon jets there is a
significant excess of events with charge $Q=0$ above the MC expectation 
with a significance of about 
$4\sigma$. A similar excess is found by OPAL, again for JETSET, but an 
even larger effect for ARIADNE\cite{ariadne} and a smaller effect for
HERWIG.\cite{herwig} The same excesses in comparison to
JETSET and ARIADNE have also been observed by ALEPH where the
effect is also shown to increase with the size of the rapidity gap and 
appears mainly at the small multiplicities $\leq4$.  
These results show the inadequacy of MC's to describe the charge
distribution of leading clusters in gluon jets. This may be taken as a hint 
towards the presence of the
octet neutralization mechanism.

The next question to ask is whether the events with a leading cluster 
of charge $Q=0$ show indeed a sign of a glueball which would be a natural
explanation of the observed excesses. The mass spectra shown by DELPHI
for quark jets are again in good agreement with the MC expectations.
The mass distribution for events with charge $Q=0$ in gluon jets 
in the mass region below
2 GeV show an excess near $f_0(980)$ and possibly in the mass region around
1400 MeV.\footnote{The significance of these excesses are still under 
experimental investigation.\cite{buschbeck}}
OPAL observes a moderate excess in the region 1.0-2.5 GeV of $2\sigma$
significance with a coarse binning whereas ALEPH did not study mass spectra 
(however, their finding of
the most sgnificant excess at low multiplicities indicates the main effect
at low mass as well).

At present one cannot claim a significant signal in the mass spectra in view
of insufficient statistics or the preliminary nature of the data. We want to
stress therefore the importance of further dedicated studies. 

If the
finding of a significant excess of $f_0(980)$ in gluon jets is confirmed
it would be a strong argument in favour of a large flavour singlet component
of this meson\cite{instanton,mo0,narison} 
which could couple to two gluons and disfavour a flavour
octet assignment.\cite{anisoct} It would also support the picture with the
correspondence of $f_0(980)$ and $\eta'$ both with $gg$ coupling but not 
being a glueball; similar effects are discussed in charmless $B$-decays.%
\cite{mo1} Also we note that such an excess is not what would be 
naturally expected from a 4 quark model for this state. 
Furthermore, there is the possibility of
the broad $0^{++}$ glueball state above 1 GeV 
which may be indicated by the above findings. 

For further improvements of
the analysis it seems important to obtain a better separation of the leading
cluster from the other particles in the event. This is not immediately
garanteed if the rapidity gap is only applied to the particles inside one
jet and not in the full event. This can be achieved if the gap is required
for the gluon jet in a frame with a symmetric configuration and the same
angle between the gluon and the two quark jets.\cite{opalsymm} Then
larger rapidity gaps can possibly be achieved with reasonable statistics
which would reduce the background from non-minimal configurations.

\section{Evidence for Scalar Glueball in Charmless B-Decay}

The interest in charmless  $B$-decays with strangeness has been stimulated
through the observation  of a large 
decay rate $B\to \eta'K$ and  $B\to \eta'X$.\cite{cleo0}
It has been suggested that these
decays, at least partially, proceed through the $b$-quark decay  $b\to sg$. 
This decay could be a source of mesons with large gluon affinity.%
\cite{soni,fritzsch,hou,dgr} In consequence, besides $\eta'$
also other gluonic states, in particular also glueballs    
could be produced in a similar way. 
 
The total rate $b\to sg$  has been calculated  perturbatively
in leading\cite{ciuchini} and next-to-leading order\cite{greub}
\begin{equation}
\text{Br} (b\to sg) =
\begin{cases}
(2-5)\times 10^{-3} & \text{in LO  (for $\mu=m_b\ldots m_b/2$)}\\
(5\pm 1)\times 10^{-3} &  \text{in NLO}
\end{cases}
\label{btosg}
\end{equation}
The energetic massless gluon in this process could turn
entirely into gluonic mesons by a nonperturbative transition
after neutralization by a second gluon.
Alternatively, colour neutralization through $q\overline q$ pairs is
possible as well (see also Sec. 4 and Fig. 1). 
This is to be distinguished from the short distance process $b\to
s\overline q q$ with virtual intermediate gluon
which has to be added to the
CKM-suppressed decays $b\to q_1\overline q_2 q_2$. These quark processes
with $s$
have been calculated and amount to branching fractions of
$\sim 2\times 10^{-3}$ each.\cite{altarelli,nierste,greub}

Recently, the  BELLE collaboration\cite{belle,belle2} has studied
charmless decays $B\to Khh$ with $h=\pi,K$ showing
a strong signal of the decay $B\to Kf_0(980)$ with
decay rate comparable to $B\to K\pi$. Similar results have been obtained by 
BaBar.\cite{babar1,babar2} 
These results show that scalar particles are easily produced and open the
possibility to identify the scalar nonet related to $f_0(980)$ and to
determine its properties.\cite{mo1}

There is another interesting feature in the decays $B^+\to
K^+\pi^+\pi^-$
and $B^+\to K^+ K^- K^+$ observed by the BELLE collaboration.\cite{belle}
The latter channel shows a broad enhancement in the $ K^+ K^-$ mass
spectrum in the region $1.0-1.7$~GeV. The flat
distribution in the Dalitz plot of
these events suggests this object to be produced with spin $J=0$. 
Its contribution 
can be parametrized as scalar state $f_X(1500)$ with mass $M=1500$ MeV
and $\Gamma=700$ MeV. More recent preliminary data\cite{belle2} 
of higher statistics indicate
a narrower substructur above the broad ``background''. In the $\pi\pi$
channel there is also some background under $f_0(980)$ which drops above
1400 MeV. 

As discussed elsewhere\cite{mo1} we consider the enhancements in $\pi\pi$
and $K\overline K$
observed by BELLE as a new 
 manifestation of the broad scalar glueball discussed above under
route II. Whereas the center of the
peak in $\pi\pi$ is closer to 1 GeV, it is shifted to higher mass in the
$K\overline K$ channel. The glueball decays with equal rates into
$u\overline u,\ d\overline d$ and $s\overline s$ and also into $gb\ gb$. 
The strange quarks produce
dominantly $K\overline K$, the nonstrange quarks produce $\pi\pi$ but also
resonances in the $4\pi$ channel (like $\rho\rho$ or $\sigma\sigma$).
We therefore interprete the drop in the $\pi\pi$ spectrum as consequence of
the opening of the $4\pi$ channel for glueball decay. An important test of
our glueball hypothesis is the verification of the expected branching
ratios; with the published data there is no contradiction with the
hypothesis.

Using the published $B$ branching ratios\cite{belle} we have 
estimated the total production rate of the scalar glueball to be
\begin{equation}
\mbox{Br} ( B^+\to gb(0^{++})+X_s) \sim 1.2\times 10^{-3}, \label{glutot}
\end{equation}
adding the gluonic part of $f_0(980)$ and $\eta'$ production  we estimate
\begin{equation}
\mbox{Br} (B^+\to gb(0^{++})+f_0+\eta'+X_s) \sim (1.5\pm0.5)\times 10^{-3}
\label{glumestot}
\end{equation}
which is of the same size as the leading order result for the process
$b\to sg$ in (\ref{btosg}) and about 1/3 of the full rate obtained in NLO.
Further gluonic contributions are expected from other glueballs, in
particular, from the parity partner $0^{-+}$ so that the total production
rate of the decay $b\to sg$ could be saturated by gluonic mesons and
gluonic production of flavour singlet mesons. 

\section{Summary}
1. Recent results from QCD calculations confirm the existence of glueballs,
the lightest one with quantum numbers $0^{++}$. The results for its mass 
vary in a range from 1000 to 1800 MeV.

2. There are different scenarios in the phenomenological interpretation of
the data: the glueball has a mass around 1500 MeV and mixes with the
isoscalar members of the nonet into observed isoscalars in the range 1300 -
1800 MeV. Alternatively, the glueball is around 1 GeV or a bit heavier
with a large width of 500-1000 MeV. In any case, there is no single
narrow resonance representing the glueball.

3. Many relevant data (decay branching ratios of $f_0$'s, production rates
and phases of $f_0$'s in various production and decay channels have often
large errors or are controversial. An improvement of this experimental
situation by better measurements is essential for further progress.
We think here in particular of improved measurements on the various $f_0$'s
(and $gb$)
in central $pp$ (double Pomeron) production, $J/\psi$ decays ($\to
f_0\omega,\ f_0\phi;\ \gamma 2\pi,\ \gamma4\pi)$ and $D$-decays ($f_0\pi$).

A crucial role is played by $f_0(980)$ which could be either a member of
the higher mass or lower mass nonet. Its properties and flavour composition
are still controversial.

4. An attractive new possibility to search for glueballs lies in the comparison
of leading clusters in gluon 
and quark jets. Also new information on 
flavour singlet mesons can be obtained ($\eta'$, $f_0(980)$?). 
First results from LEP
show an additional neutral component in gluon jets not expected from
standard MC's. The origin of this excess has to be understood
in terms of hadronic states.  Such studies are possible at HERA.

5. Charmless $B$-decays are an attractive source of glueballs and other 
gluonic mesons from the decay $b\to sg$ as well. 
Recent results have been interpreted in terms of a broad scalar glueball
decaying into $K\bar K$ in the mass range 1000-1600 MeV.

The recent results from $B$ decays and gluon jets are quite promising.
They allow to contrast quark and gluon structures in the same experiment.
As we hope, there is a large potential to establish the light 
scalar $q\bar q$ nonet
with its mixing and the lightest scalar glueball.

%%%%%%%%%%%%%%%%%%%%%%%%%%%%%%%%%%%%%%%%%%%%%%%%%%%%%%%%%%%%%%%%%%%%%%%

%\input{refs}
%*** File 'refs'

\end{document}